\documentclass{svproc}

\usepackage{xcolor}
\usepackage{graphicx,amsmath}

\usepackage{url}

\begin{document}
\mainmatter              
\title{Towards a parameter-free analysis of the QCD chiral phase transition and its universal critical behavior}
\titlerunning{Universality of chiral phase transition}  
\author{Sabarnya Mitra\inst{1}
\and Frithjof Karsch\inst{1}}

\institute{Fakult\"at f\"ur Physik, Universit\"at Bielefeld, D-33615 Bielefeld,
Germany \\
\email{smitra@physik.uni-bielefeld.de}, \,
\email{karsch@physik.uni-bielefeld.de}
}

\maketitle

\newcommand{\muB}{\mu_B}

\newcommand{\barechiralcondinlattice}{\left \langle \Bar{\psi}\,\psi \right \rangle}
\newcommand{\barechiralcond}{M_\ell}
\newcommand{\barechiralsusc}{\chi_\ell}
\newcommand{\barechiralcondreg}{M_{\ell, \text{reg}}}
\newcommand{\barechiralsuscreg}{\chi_{\ell, \text{reg}}}
\newcommand{\impchiralcondreg}{M_{\text{reg}}}
\newcommand{\condsclfunc}{f_G}
\newcommand{\suscsclfunc}{f_{\chi}}
\newcommand{\imprsclfunc}{f_{G\chi}}

\begin{abstract}

To quantify the universal properties of chiral phase transition in (2+1)-flavor QCD,
we use an improved, renormalized order parameter
for the chiral symmetry breaking. We construct  
ratios of this divergence-free order parameter
from its values for different pairs of light quark masses. 
From this, we determine in a parameter-independent manner, the chiral phase transition temperature $T_c$ and the associated critical exponent $\delta$ of the universality class. 
We present first numerical results of these
calculations on $N_\tau=8$ lattices, with staggered fermions. 

\end{abstract}

\section{Introduction}

With the order of Quantum Chromodynamics (QCD) chiral phase transition and associated universal properties still controversially poised \cite{Pisarski:1983ms,Pelissetto:2013hqa}, the extent of its influence on respective physical QCD thermodynamics and QCD axial anomaly constitute questions of profound importance. In this work, we try addressing these challenges by obtaining a first principle numerical analysis of the universal, critical behaviour of (2+1)-flavor QCD on lattice, with physical values of strange quark masses. Besides extending earlier studies of magnetic QCD equation of state \cite{Ejiri:2009ac,Bazavov:2011nk,Ding:2024sux}, this work also attempts to quantify the relevant scaling regimes for various physical observables \cite{Kotov:2021rah,Braun:2023qak}, which is pivotal for the interpretation of heavy-ion collision data in terms of
chiral critical behavior.

\section{Theory} 
 
\begin{figure}[h!]
\begin{center}
\includegraphics[width=0.41\textwidth]{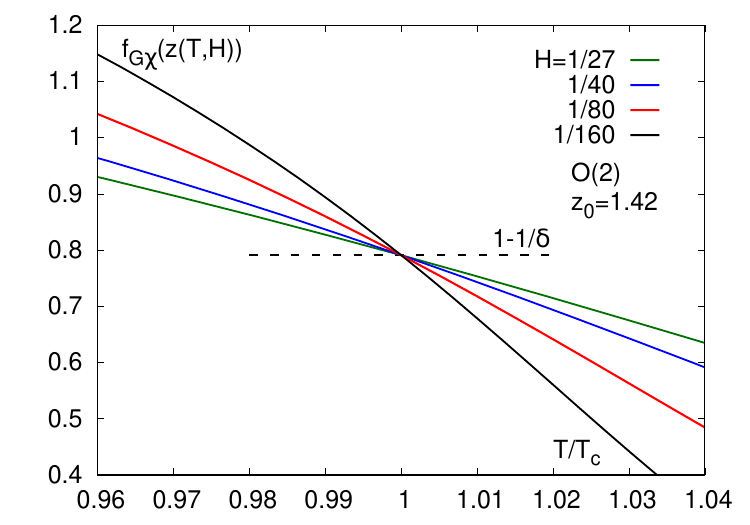}
\hspace{.7cm}
\includegraphics[width=0.41\textwidth]{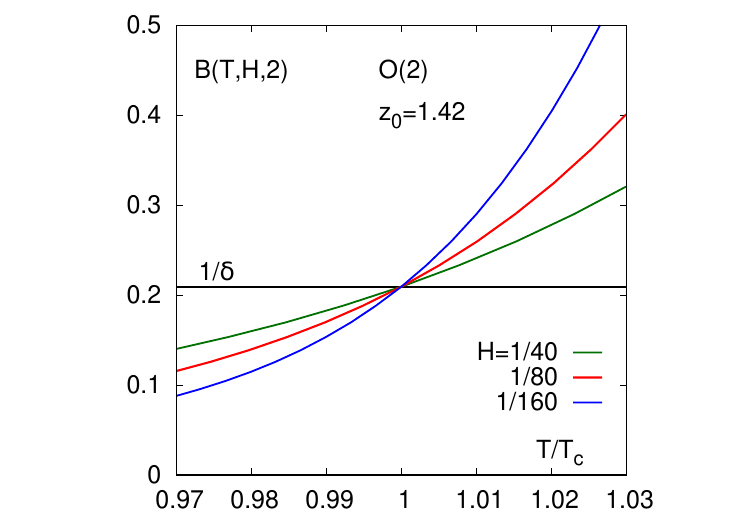}
\caption{ 
%For different values of $H$, 
{\it Left:} Scaling function $f_{G\chi}$ vs $T/T_c$ for different values of $H$. Shown is $f_{G\chi}$
for the $3$-d, O(2) universality class
using the non-universal parameter
$z_0=1.42$ appropriate for (2+1)-flavor QCD in calculations with the HISQ action
on discrete lattices of temporal extent $N_\tau=8$.
{\it Right:} Same as left hand figure but for the observable $B(T,H,2)$.
}
\label{fig:fGchi and B}
\end{center}
\end{figure}

%\subsection{Theoretical background}
 With light quark mass $m_\ell$, the chief ingredients in this work are the $2$-flavor light quark chiral condensate, $\barechiralcond$ and the corresponding chiral susceptibility $\barechiralsusc$, 
  \begin{equation}
      \barechiralcond = \frac{m_s}{f_K^4} \barechiralcondinlattice, \hspace{1cm}
      \barechiralsusc = m_s \,\frac{\partial \barechiralcond}{\partial m_\ell} \; ,
      \label{basics}
  \end{equation}
  where the strange quark mass $m_s$ and kaon decay constant $f_K$ are normalising factors.
To attain a well-defined order parameter $M$ in the continuum limit for studying the chiral symmetry breaking, the ultraviolet divergent contributions to $M_\ell$
  are eliminated by deducting from it, a suitable
  fraction of $\chi_\ell$ :

 \begin{equation}
     M(T,H) = \barechiralcond (T,H) - H\,\barechiralsusc(T,H)\; ,
     \label{eq:M}
 \end{equation}
where $H\equiv m_\ell/m_s$. 
Thus in chiral limit, $M$ in Eq.\eqref{eq:M} receives in the leading order, $\mathcal{O}(H^3)$ light quark mass corrections. 
This improved order parameter
has been introduced in \cite{Unger:2010wcq} and used previously to probe critical behavior in lattice QCD \cite{Ding:2024sux,Kotov:2021rah,Dini:2021hug}. 
Close to the chiral critical point $(T=T_c,\,H=0)$, the $T$ and $m_\ell$ dependence of $M$ is largely governed by the scaling function $f_{G\chi}(z)$ like,
\begin{equation}
    M(T,H) = h^{1/\delta} f_{G\chi}(z) + M_{sub-lead}(T,H) \; ,
    \label{eq:M-scaling}
\end{equation}
where $f_{G\chi}(z)=f_G(z)-f_\chi(z)$. The scaling functions $f_G,\,f_\chi$ control the quantities $M_\ell,\,\chi_\ell$ respectively in a manner, similar to Eq.\eqref{eq:M-scaling}. 
Corrections to the leading, so-called singular part of the order parameter arise
from {\it i)} non-analytic, universal corrections-to-scaling and {\it ii)} analytic, non-universal regular terms.
They are summarized in $M_{sub-lead}$
and have a leading $H$-dependence of 
higher order in $H$.
The scaling variable $z$, used in Eq.\eqref{eq:M-scaling} is given as, 
 \begin{equation}
     z = t\,h^{-1/\beta\delta}, \hspace{.6cm} \text{where} \hspace{.4cm} t = \frac{1}{t_0}\,\left[\frac{T}{T_c}-1\right], \hspace{.4cm} h = \frac{H}{h_0}\; ,
     \label{eq:scaling variable}
 \end{equation}
 with non-universal constants $t_0,h_0$, and $\beta, \delta$ being critical exponents of the related universality class.
 Using properties of $f_G$ and $f_\chi$\,\cite{Karsch:2023pga}, one can easily find $f_{G\chi}(0)=1-1/\delta$. When plotted as a function of $T$ for different values
of $H$, this manifests a unique intersection point at $z=0$, which directly provides values of $T_c$ and $\delta$ (see left plot of Fig. \ref{fig:fGchi and B}). Exploiting this property, we construct the re-scaled order parameter (resc. in $H$)
 \begin{equation}
      M(T,H)/H^{1/\delta} = h_0^{-1/\delta}
     f_{G\chi}(z) + H^{-1/\delta} M_{sub-lead}(T,H)\; ,
     \label{eq:M resc}
 \end{equation}
 and the quantity $B$ from the ratio of $M$ for two different values of $H$ as, 
  \begin{eqnarray}
     B(T,H,c) &=& \frac{\ln R(T,H,c)}{\ln (c)}\,,\; \,\,\,\,\text{with}\,\,\, R(T, H, c) = \frac{M(T,cH)}{M(T,H)} \; .
 \label{M-ratiolog}
 \end{eqnarray}

 \noindent
 In chiral limit, $H^{-1/\delta}M_{sub-lead} \to 0$ which automatically gives $B(T_c,0,c) = 1/\delta$.

\section{Results} 

\begin{figure*}
\centering
%\hspace{-1cm}
\includegraphics[width=0.53\textwidth]{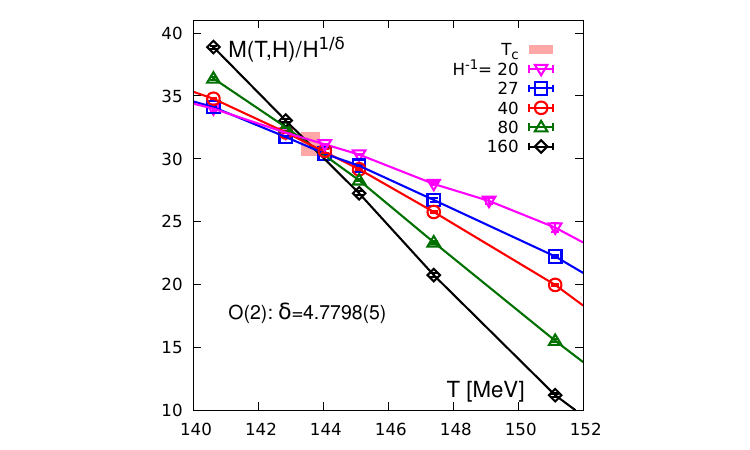}
 \hspace{-1cm}
\includegraphics[width=0.53\textwidth]{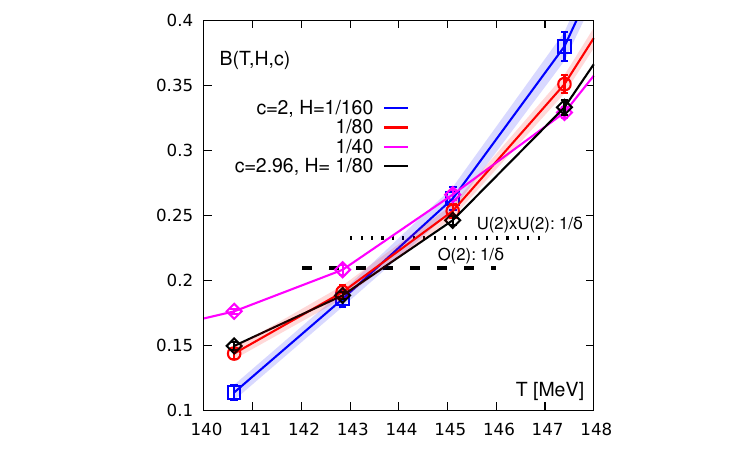}
\caption{{\it Left:} The rescaled order parameter, $M(T,H)/H^{1/\delta}$ vs. $T$ for several $H$ values. Shown are results
obtained in lattice QCD calculations with
the HISQ action on lattices with temporal
extent $N_\tau=8$.
{\it Right:} The observable $B(T,H,c)$ vs. $T$ for different values of $c$ and $H$. The dashed and dotted lines give the values for $1/\delta$ in $O(2)$ and $U(2)\times U(2)$ \cite{Pelissetto:2013hqa} universality classes, respectively.
}
\label{fig:resc-M-and-B}
\end{figure*}

Here in this work, we have used HISQ and Symanzik-improved gauge actions for all the lattice computations. The relevant $T$ and $H$ data are taken from \cite{Ding:2024sux,HotQCD:2019xnw}. In Fig.~\ref{fig:resc-M-and-B}~(left), we show results for 
$M/H^{1/\delta}$ using $\delta=4.7798$, its value in the $3$-dimensional $O(2)$ universality class as HISQ preserve only the $O(2)$ part out of full $O(4)$ symmetry. A unique intersection point is apparent from Fig.~\ref{fig:resc-M-and-B}~(left). Its location agrees well
with the previous determination of $T_c$ 
{\it i.e.} $T_c^{N_\tau=8}=143.7(2)$~MeV
\cite{Ding:2024sux}, and $h_0^{-1/\delta}$ (Eq.\eqref{eq:M resc})\,; $h_0^{-1/\delta}=39.2(4)$ \cite{Ding:2024sux}.
For a parameter-independent analysis, we evaluate
$B(T,H,c)$  
as shown in Fig. \ref{fig:resc-M-and-B}. This plot also features a distinct intersection point, giving $T_c,\,\delta$ but only for small enough $m_\ell \leq m_s/40$, thereby indicating that contributions from $M_{sub-lead}$ may become significant for $H > 1/40$, at least for the present working level of precision.

\section{Conclusion and Outlook}

We have demonstrated in this work, that as a function of $T$, the re-scaled order parameter $M$ exhibits a unique intersection point for different values of light quark mass $m_\ell$, or equivalently $H$ for small enough values of $m_\ell$ where the sub-leading contributions are negligible. We also showed that evaluating its ratios allows determining $T_c$ and critical exponent $\delta$, without having the need to make any apriori assumption about the universality class. However, we still need to improve computational precision by incorporating more $T,H$ data points close to critical point, besides checking the finite-volume effects and taking the continuum limit. This is to estimate 
$\delta$ with accuracy good enough to distinguish between $U(2)\times U(2)$ and $O(4)$ universality classes. 

\section*{Acknowledgments}
This work was supported by the Deutsche Forschungsgemeinschaft
(DFG, German Research Foundation) Proj. No. 315477589-TRR 211. Numerical computations
have been performed on the GPU-cluster at Bielefeld University.

\end{document}